\documentclass[usenatbib]{mn2e}
\usepackage{graphicx,amsmath,multirow,amssymb}
\usepackage{subfig}
\usepackage{natbib}
\usepackage[usenames,dvipsnames]{color}

\newcommand{\comment}[1]{}

\def\simgt{\lower.5ex\hbox{$\; \buildrel > \over \sim \;$}}
\def\simlt{\lower.5ex\hbox{$\; \buildrel < \over \sim \;$}}
\newcommand{\Mgvq}{\ensuremath{{^{24}{\rm Mg}}}}
\newcommand{\Mgvc}{\ensuremath{{^{25}{\rm Mg}}}}
\newcommand{\Mgvs}{\ensuremath{{^{26}{\rm Mg}}}}
\newcommand{\Msun}{\ensuremath{{\rm M}_{\sun}}}

\newcommand\aap{{A\&A}}%
\newcommand\mnras{{MNRAS}}%
\title[Magnesium processing in HBB - AGBs]{Magnesium isotopes: a tool to understand
self-enrichment in Globular Clusters}
\author[Ventura et al.]{P. Ventura$^1$, F. D'Antona$^1$, G. Imbriani$^2$,  
M. Di Criscienzo$^1$, F. Dell'Agli$^{3,4}$, 
\newauthor
M. Tailo $^{5}$ \\ 
$^{1}$INAF -- Osservatorio Astronomico di Roma, Via Frascati 33, 00078, Monte Porzio Catone (RM), Italy \\
$^{2}$Dipartimento di Scienze Fisiche, Universit\'a di Napoli Federico II, Italy \\
$^{3}$Instituto de Astrof\'{\i}sica de Canarias, E-38205 La Laguna, Tenerife, Spain \\
$^{4}$Departamento de Astrof\'{\i}sica, Universidad de La Laguna (ULL), E-38206 La Laguna, Tenerife, Spain\\
$^{5}$Dipartimento di Fisica, Universit\'a degli Studi di Cagliari, SP Monserrato-Sestu km 0.7, I-09042 Monserrato, Italy
}

\begin{document}

\date{Accepted, Received; in original form }

\pagerange{\pageref{firstpage}--\pageref{lastpage}} \pubyear{2012}

\maketitle

\label{firstpage}

\begin{abstract}
A critical issue in the asymptotic giant branch (AGB) self-enrichment scenario for 
the formation of multiple populations in Globular Clusters (GCs) is the inability to 
reproduce the magnesium isotopic ratios, despite the model in principle can account for 
the depletion of magnesium. 
In this work we analyze how the uncertainties on the various p-capture cross
sections affect the results related to the magnesium content of the ejecta of AGB stars.
The observed distribution of the magnesium isotopes and of the overall Mg-Al trend in M13 
and NGC 6752 are successfully reproduced when the proton-capture rate by $^{25}$Mg
at the temperatures $\sim 100$ MK, in particular the $^{25}{\rm Mg}(p,\gamma)^{26}{\rm Al^{m}}$ 
channel, is enhanced by a factor $\sim 3$ with respect to the most recent experimental 
determinations. This assumption also allows to reproduce the full extent of the Mg spread 
and the Mg-Si anticorrelation observed in NGC 2419. 
The uncertainties in the rate of the $^{25}{\rm Mg}(p,\gamma)^{26}{\rm Al^{m}}$ 
reaction at the temperatures of interest here leave space for our assumption and we
suggest that new experimental measurements are needed to settle this problem.
We also discuss the competitive model based on the super massive
star nucleosynthesis.
\end{abstract}


\section{Introduction}
The research focused on Globular Clusters received a boost in the last decades,
owing to results from high resolution spectroscopy, photometry and spectrophotometry, which have
challenged the traditional paradigma that GCs are simple stellar populations.

On the spectroscopic side, practically all the GCs of the
Milky way so far examined present star-to-star variations, which trace well defined abundance patterns,
such as the C-N and O-Na anti-correlations, whose extension varies from cluster to cluster \citep{gratton12}. 
These abundance variations were observed also at the surface of unevolved stars \citep[e.g.][]{gratton01};
for these stars, unlike red giants, the effects of any possible `in situ' production mechanism
can be disregarded. This discovery lead to the conclusion that in GCs a variety of generations
of stars coexist, each characterized by a different chemical composition. 

On the photometric side, a great step forward towards the identification of multiple stellar
generations was taken when the presence of a ``blue main sequence" was discovered
in $\omega$ Centauri \citep{bedin04} and in NGC 2808 \citep{piotto07}, 
requiring the presence of a group of stars formed with a helium content much larger 
than the standard Big Bang abundance. The blue main sequences confirmed early predictions, based on the analysis of the complex morphology of the horizontal branch (HB) of some GCs 
\citep{dantona02, dantona04}. In NGC 2808, three groups of stars differing in their helium 
were shown to coexist, with helium mass fractions ranging from the primordial, $Y \simeq 0.25$ 
up to $Y \simlt 0.40$ \citep{dantona05, piotto07}. Conversely, in other GCs, such as 47 Tuc, 
only a very modest spread of helium is required to reproduce the width of the main 
sequence (MS) and the morphology of the HB \citep{marcella10}.

The Hubble Space Telescope (HST) UV Legacy Survey \citep{piotto15} has recently 
exploited the sensitivity of UV photometric observations to different molecular bands, to 
disentangle the stellar populations in 57 GCs. This project allowed to sort out
the different stellar components present in each GC, providing a valuable tool to 
understand how multiple populations formed and evolved in Galactic GCs.

The combination of the spectroscopic and spectro--photometric evidences collected so far indicate
that in GCs, after the formation of the first generation (FG), a second generation (SG)
of stars formed, from gas contaminated by p-capture nucleosynthesis. 
Several star formation events may contribute to different SG groups \citep{milone17}.

Until now 4 main classes of polluters have been proposed, namely (in order of decreasing mass): 
super-massive main sequence stars \citep{denissenkov14}; fast rotating massive stars 
\citep{decressin07, krause13}, massive binaries \citep{demink09} and massive AGB stars \citep{dercole08}.
In all models, the proton capture reactions, producing the peculiar abundances found in the SG stars, 
occur in the convective hydrogen burning core; the only exception is the AGB 
case, where the site of nucleosynthesis is the bottom of the deep convective envelope 
(hot bottom burning, HBB). 
All the proposed scenarios have been subject to heavy criticism, in particular concerning 
the mass budget issue \citep[see, e.g.][]{renzini15}, so that the most recent proposal is that the puzzle of 
multiple populations remains unsolved, hence alternative theories are needed \citep{bastianlardo2017}.
Nevertheless, the discrepancies between models and observations are not of the same level. 
In this work, we limit the comparison to the models in which nuclear processing is able to trace the most extreme chemical patterns 
so far found in the SG stars. Only the supermassive star model and the AGB model can, at 
least qualitatively, provide the proton captures on magnesium nuclei required to explain 
the Mg--Al anticorrelation found in a few clusters, so here we focus on
the predictions of these two models.

Sufficiently large samples of data 
on the Mg-Al anti-correlation have been acquired only in recent times. In clusters where 
it is present, the Mg--Al data provide more information on the source of pollution than 
the standard feature of multiple populations, the O--Na  anticorrelation, for the 
following reasons: a) proton captures on Mg nuclei are activated at higher temperatures than 
those on  $^{16}$O and $^{22}$Ne nuclei (the latter reaction leading to the synthesis of Na); 
b) Mg and Al abundances measured in giant stars definitely reflect the initial chemistry, as
in the low mass stars evolving today in GCs, these elements are not subject to any nuclear 
processing, not even in the central regions. In addition Mg is a key element for 
any scenario, because its abundance can only decrease as a result of evolution 
(no production is allowed), so it is a direct signature of nuclear processing in the
polluters\footnote{For instance, in the AGB scenario, the abundance of sodium in SG stars depends 
in great part also on the effect of the second dredge up, which brings to the envelope both 
sodium and the neon isotopes processed in the interior. Fast p-captures on the dredged up 
neon contribute to increase sodium during the first phases of HBB \citep[e.g.][]{vd06}. This is often forgotten 
in the simplified explanations attributing only the helium abundance of AGBs to the second 
dredge up.}. 

The most recent results have confirmed the early discovery that the isotopic 
magnesium ratios in NGC\,6752  \citep{yong03} do not agree with the abundance ratios 
predicted in the HBB models computed so far \citep{ventura09}.
On the other hand, the observed abundance ratios \Mgvc/\Mgvq\ and \Mgvs/\Mgvq\ are more 
easily achieved in the convective cores of those supermassive stars that burn hydrogen at 
temperatures $\simeq$75\,MK \citep{denissenkov14, denissenkov15}. In principle, this
suggests that the interiors of such supermassive stars are the most plausible site for the 
Mg nucleosynthesis producing the chemical patterns observed in GC stars. 
We will discuss later on the difficulties of supermassive star model.

The paper is structured as follows. We discuss the relevant observations in 
Section\,2, and the Mg--Al nucleosynthesis and p-capture cross sections in Section\,3.
We examine in detail the Mg-related nucleosynthesis occurring in the models of 
massive AGB stars for which we have produced yields in the recent literature. 

In Section\,4 the theoretical predictions, based on  the chemical composition of updated 
AGB yields, are compared with the observations of the Mg-Al trends and with the magnesium isotopic 
ratios of stars belonging to GCs of different mass and metallicity. 
We focus on the observations of the magnesium 
isotopes in M13 and NGC 6752, in the attempt to reproduce the relative fractions
of $^{25}$Mg and $^{26}$Mg with respect to the total magnesium and the extension of
the Mg--Al pattern. 
We consider the nucleosynthesis resulting from the nuclear reaction rates currently 
available (the LUNA compilation, Strieder et al. 2012) and we study which variations of 
these rates may produce  full agreement with the the data.  We also extend the computation to 
the chemistry of NCG 2419, a cluster where the abundances of the individual Mg isotopes 
are not available, but  hosting the largest observed spread (by a factor $\sim 10$) in 
the overall magnesium abundance. 

In Section 5 we perform a global comparison between the model predictions and the 
data, and show that these AGB models, in the framework of the dilution model, 
are compatible both with the depletion 
of magnesium in the SG in clusters of different metallicity, and with the increase 
in helium accounting for the MS width and the HB morphology.  We also discuss 
the results of the supermassive star models. Finally, in Section 6 we discuss that the 
suggested increase in the $^{25}$Mg proton capture rate is plausible, and make a plea for a 
new experimental determination.

\section{Observed Mg and Al abundances in GCs}
Early detections of star-to-star variations in Al line strength 
\citep{norris81, norris83, cottrell81} were confirmed as being due to Al abundance 
variations \citep{drake92, brown92, norris95}.
The presence of a  Mg--spread among GC stars was set by the pioneering investigations
by \citet{shetrone96a}, who detected star-to-star differences in the Mg content of giant stars 
in M13, and by \citet{king98}, who measured extremely low Mg abundances in stars populating the
sub giant branch of M92. Later works detected hints of a possible Mg-Al anti-correlations 
in M3 \citep{cavallo00, johnson05}, M13 \citep{cavallo00, sneden04, cohen05} and
NGC 6752 \citep{gratton01}.

A robust and quantitative confirmation of  the Mg--Al anti-correlation in some GCs came with
the works by \citet{carretta09} and \citet{meszaros15}. Some clusters also display the presence 
of a Mg--Si direct correlation. Additional, recent data on Mg--Al were presented 
by \citet{carretta12a, carretta12b, carretta14, carretta15a, carretta15b}, \citet{gruyters14}. 

\begin{table*}
\caption{The maximum variation (SG-FG), $\delta(X)$, in magnesium, aluminium and silicon, 
measured in stars of different GCs, based on the number of stars N, indicated in col. 7. 
The helium maximum variation, $\delta Y$ (col. 8), is based on models reproducing 
the MS and/or the morphology of the HB.}                                       
\begin{tabular}{c c c r c c c l}        
\hline       
name & & [Fe/H]& $\delta [Mg/Fe]$ & $\delta [Al/Fe]$ & $\delta [Si/Fe]$ & N & $\delta Y$ \\
\hline  
 NGC 7078  &  M15   & -2.37 &    $-0.50^1$ & 1.10 & 0.40 &   23  &  $0.07^8$     \\
 NGC 7099  &  M30   & -2.34 &    $ 0.00^2$ & 1.20 & 0.20 &   10  &  $0.02^9$     \\
 NGC 6341  &  M92   & -2.31 &    $-0.60^1$ & 1.10 & 0.30 &   47  &               \\
 NGC 4590  &  M68   & -2.26 &    $-0.10^2$ & 1.00 & 0.10 &   13  &               \\
 NGC 2419  &        & -2.10 &    $-1.00^7$ &  -   & 0.30 &   13  &  $0.12^{10}$  \\
 NGC 5024  &  M53   & -2.10 &    $-0.20^1$ & 1.10 & 0.00 &   16  &              \\
 NGC 4833  &        & -2.02 &    $-0.50^4$ & 1.20 & 0.20 &  ~50  &              \\
 NGC 5466  &        & -1.98 &    $0.00^1$  & 0.50 & 0.00 &    3  &              \\
 NGC 6809  &  M55   & -1.93 &    $-0.30^2$ & 1.00 & 0.00 &   14  &              \\
 NGC 6093  &  M80   & -1.79 &    $-0.30^5$ & 1.30 & 0.10 &   13  &              \\
 NGC 7089  &  M2    & -1.65 &    $-0.15^1$ & 1.20 & 0.00 &   18  &              \\ 
 NGC 1904  &  M79   & -1.57 &    $0.00^2$  & 1.00 & 0.00 &   10  &              \\
 NGC 6254  &  M10   & -1.57 &    $0.00^2$  & 1.00 & 0.10 &   14  &              \\
 NGC 6752  &        & -1.55 &    $-0.15^2$ & 1.20 & 0.20 &   14  &  $0.07^{16}$     \\
 NGC 6205  &  M13   & -1.53 &    $-0.40^1$ & 1.40 & 0.00 &   81  &  $0.10^{11}$  \\
 NGC 3201  &        & -1.51 &    $0.00^2$  & 1.00 & 0.00 &   13  &              \\
 NGC 5272  &  M3    & -1.50 &    $-0.18^1$ & 1.00 & 0.00 &   59  &  $0.05^{12}$  \\
 NGC 6218  &  M12   & -1.33 &    $0.00^2$  & 0.50 & 0.00 &   11  &              \\
 NGC 288   &        & -1.30 &    $0.00^2$  & 0.00 & 0.00 &   10  &  $0.015^{13} $\\
 NGC 5904  &  M5    & -1.29 &    $0.00^1$  & 1.00 & 0.00 &  102  &              \\ 
 NGC 6121  &  M4    & -1.17 &    $0.00^2$  & 0.00 & 0.10 &   14  &              \\
 NGC 1851  &        & -1.15 &    $-0.08^3$ & 0.80 & 0.05 &   60  &              \\
 NGC 2808  &        & -1.12 &    $-0.40^6$ & 0.00 & 0.10 &  139  &  $0.13^{14}$  \\
 NGC 6171  &  M107  & -1.02 &    $0.00^1$  & 0.70 & 0.00 &   12  &              \\
 NGC 6838  &  M71   & -0.83 &    $0.00^2$  & 0.70 & 0.10 &   12  &              \\
 NGC 104   &  47Tuc & -0.77 &    $0.00^2$  & 1.00 & 0.00 &   11  &  $0.03^{15}$  \\
\hline       
\label{tabmg}
\end{tabular}

{The references of the data on the various clusters are the following: 1 - Meszaros et al. 
(2015); 2 - Carretta et al. (2009); 3 - Carretta et al. (2012b); 4 - Carretta et al. (2014);
5 - Carretta et al. (2015); 6 - Carretta (2015); 7 - Carretta et al. (2014); 8 - Milone et 
al. (2013); Mucciarelli et al. (2014); 10 - Di Criscienzo et al. (2015); 11 -
D'Antona \& Caloi (2008); 12 - Caloi \& D'Antona (2008); 13 - Piotto et al. (2013);
14 - D'Antona \& Caloi (2013); 15 - Di Criscienzo et al. (2010); 16 - Tailo et al., in preparation
}
\end{table*}

We present in Table \ref{tabmg} a compilation of the most recent results, for the GCs 
with the largest statistics, reporting the extent of the magnesium, 
aluminium and silicon spreads detected. We also show, when available, the initial helium 
range for the stars in the same cluster. The latter quantity is not directly measured, 
but is derived from theoretical models, either applied to describe the width of the MS or the 
morphology of the HB. The helium range is a key indicator  of the modality with which 
formation of SG stars took place and monitors the possible  dilution with pristine gas; 
we will return to this point in section \ref{disc}.

The largest differences between the magnesium measured in FG and SG stars, generally of 
the order of $\delta$[Mg/Fe]$\sim $--0.5, are observed in metal poor clusters, with 
[Fe/H]$\simlt$--2;\footnote{here and throughout the paper we indicate with $\delta$[X/Fe] 
the difference between the abundances of element X measured in SG stars, to which we 
subtract the corresponding mass fraction of FG stars}
in this metallicity domain we find a variety of situations, ranging from the cluster
M\,30, where no Mg--spread is observed and the helium spread is estimated to be within
$\delta$Y$\simeq$0.02 \citep{mucciarelli14}, to NGC 2419, a cluster harbouring a population
greatly enriched in helium \citep{marcella15}, where an extremely large
magnesium spread, $\delta$[Mg/Fe] $\sim$--1, is observed \citep{cohen12, mucciarelli12}. 
In the metallicity range --2$<$[Fe/H]$<$--1,  the only two GCs exhibiting a significant 
magnesium spread, of the order of $\delta$[Mg/Fe]$\sim -0.4$, are M\,13 and NGC\,2808, 
both commonly believed
to harbor a stellar population with a very extreme chemistry, also significantly
enriched in helium; conversely, no spread is observed in NGC 288, a cluster 
with a metallicity ([Fe/H]=--1.3) lower than NGC\,2808 ([Fe/H]=--1.1), for which 
the helium spread was estimated to be $\delta$Y = 0.015 by \citet{milone14}. Finally, 
no magnesium spread has so far been detected in GCs with [Fe/H]$>-1$.

A further information on the conditions at which the
gas from which SG stars in GCs formed can be deduced by the relative distribution
of the three magnesium isotopes, which are extremely sensitive to the details of
the nucleosynthesis experienced. On this regard, \citet{yong03} studied the distribution
of magnesium isotopes in giant stars of NGC 6752: the main finding was an overall
magnesium spread $\delta$[Mg/Fe]$\sim$ --0.2, with the distribution among the various
isotopes ranging from $^{24}$Mg:$^{25}$Mg:$^{26}$Mg=80:10:10, for FG stars, to
60:10:30, for SG stars. \citet{shetrone96b} presented Mg data of 7 giant stars in 
M13, four out of which exhibited super-solar $(^{25}Mg+^{26}Mg)/^{24}Mg$ ratios.
These results were confirmed and completed in a following study by \citet{yong06},
who found that in the most contaminated M13 stars, with the lowest magnesium and the 
largest aluminium, the distribution among the magnesium isotopes is 48:13:40, against the 
corresponding ratios detected in FG stars, found to be 78:11:11. 
Finally, \citet{dacosta13} analysed two stars in M4, finding no appreciable spread in the
distribution of the magnesium isotopes, $^{24}$Mg:$^{25}$Mg:$^{26}$Mg$\simeq$ 80:06:15
in both stars.

\section{Mg-Al nucleosynthesis}
\subsection{The p-capture reaction chain}
The details of proton-capture nucleosynthesis by magnesium nuclei was discussed in detail 
by \citet{arnould99}, \citet{boeltzig16} and \citet{iliadis11}, where the interested reader 
can find a thorough description of the relative importance of the different reaction 
channels. 

When $\rm T\simeq 60 - 100\,$MK the so called MgAl cycle starts. During the last 
decade the nuclear reactions involved in this cycle received increasing attention.
In particular, the radiative captures involving $^{24,25,26}$Mg isotopes and $^{26}$Al have 
been deeply studied. The MgAl cycle is initiated by the 
$^{24}{\rm Mg}(p,\gamma)^{25}{\rm Al}$ reaction. $^{25}{\rm Al}$ decays into 
$^{25}{\rm Mg}$, which may capture another proton yielding to either $^{26}{\rm Al^{g}}$ 
ground state, or $^{26}{\rm Al^{m}}$, metastable at $E_X=228$ keV. $^{26}{\rm Al^{g}}$ 
decays via $\beta^+$ with a half life of 0.7\,My into the first excited state of $^{26}$Mg 
with a subsequent $\gamma$-ray emission. On the contrary $^{26}{\rm Al}^m$ $\beta^+$ decays, 
with a short half life of $\tau_{1/2}^m$ = 6.3\,s, exclusively to the ground state of $^{26}$Mg.
Because of this nuclear pattern the feeding probabilities of ground state and metastable 
isotopes have particular relevance for the understanding of magnesium nucleosynthesis in 
stars. The last two reactions to consider are $^{26}{\rm Al}(p,\gamma)^{27}{\rm Si}$ and 
$^{26}{\rm Mg}(p,\gamma)^{27}{\rm Al}$.

The slowest processes in the cycle are $^{24}{\rm Mg}(p,\gamma)^{25}{\rm Al}$ and 
$^{25}{\rm Mg}(p,\gamma)^{26}{\rm Al}$, which mainly determine the time scale of the
Mg nucleosynthesis.
From the nuclear point of view the $^{24}{\rm Mg}(p,\gamma)^{25}{\rm Al}$ reaction at 
astrophysical energies has a contribution by a low-energy resonance and a strong direct 
capture component dominates the resonance contribution \citep{trautvetter75, powell99, 
iliadis10}. The present uncertainty of the reaction rate at the temperature of interest 
for the Mg nucleosynthesis is about 20\%.

The rate of the  $^{25}{\rm Mg}(p,\gamma)^{26}{\rm Al}$ reaction is characterized by 
several narrow resonances (Boeltzig et al. 2016, and references therein). In particular, 
the resonance at 92\,keV is thought to be the most important for temperatures ranging from 
50 to 120\,MK. Recently, this resonance has been directly studied by the LUNA experiment 
\citep{limata10, strieder12}, providing an update of the rates of 
$^{25}{\rm Mg}(p,\gamma)^{26}{\rm Al^{g}}$ and $^{25}{\rm Mg}(p,\gamma)^{26}{\rm Al^{m}}$. 
In particular, for 50$<$T$<$150\,MK, the rate of the $^{25}{\rm Mg}(p,\gamma)^{26}{\rm Al^{m}}$ 
production was found to be 4 times higher, while the $^{25}{\rm Mg}(p,\gamma)^{26}{\rm Al^{g}}$ 
20\% higher than previously assumed. At  $T=100\,$MK the revised total reaction rate 
\citep{straniero13} 
was determined to be a factor of 2 higher than previous determinations. The uncertainty is 
larger than for the $^{24}{\rm Mg}(p,\gamma)^{25}{\rm Al}$ reaction, since there is no 
information about the $^{26}{\rm Al}$ level corresponding to this resonance. 

The measurements performed at LUNA suggest a stronger feeding of $^{26}{\rm Al}$ states 
that predominantly decay to the isomeric state, reducing the ground state fraction 
\citep{strieder12, straniero13}.

\begin{figure}
\resizebox{1.\hsize}{!}{\includegraphics{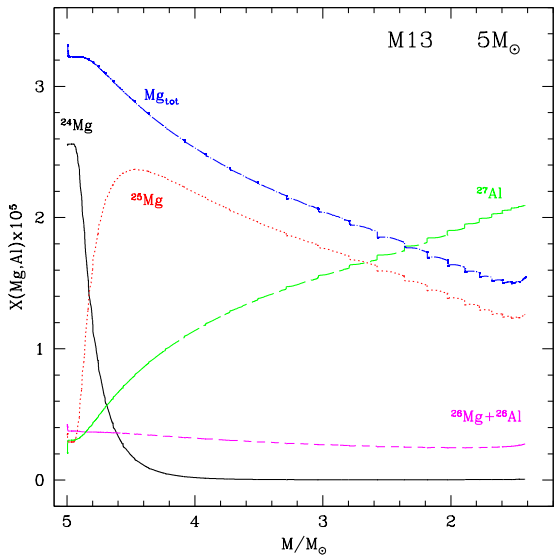}}
\vskip-60pt
\caption{The evolution of the surface mass fractions of $^{24}$Mg (black, solid line), 
$^{25}$Mg (red, dotted), ($^{26}$Mg+$^{26}$Al) (magenta, dashed), $^{27}$Al 
(long-dashed, green) and total magnesium (dotted-dashed, blue) in a model of initial
mass $5~M_{\odot}$, with the same chemical composition of stars belonging to the
FG of M13. We report the (current) mass of the star on the abscissa, to have a better
idea of the chemistry of the ejecta.}
\label{f5msun}
\end{figure}

\subsection{P--capture reactions on Mg isotopes in massive AGB envelopes}
A series of studies showed that magnesium burning can be easily achieved in
low-metallicity, massive AGB stars \citep{ventura11a, ventura11b, ventura13}.
The modality with which Mg-poor ejecta are produced by AGB stars is related to the
ignition of HBB, a physical mechanism by which
the base of the external envelope of $M \geq 4~M_{\odot}$ stars reaches temperatures 
above $\sim 30$MK, sufficiently hot to start proton-capture nucleosynthesis 
\citep{renzini81, blocker91}.
The same studies showed that the activation of magnesium burning is extremely
sensitive to the metallicity, much more than other channels, such as C-N and
Ne-Na burning: this is due to the higher temperatures required to start $^{24}$Mg
burning in AGB stars \citep{izzard07}, of the order of $T\sim 90$ MK, and to the 
higher efficiency of HBB in low-metallicity AGB stars 
\citep{ventura08, ventura09, ventura13}.

\begin{figure}
\resizebox{1.\hsize}{!}{\includegraphics{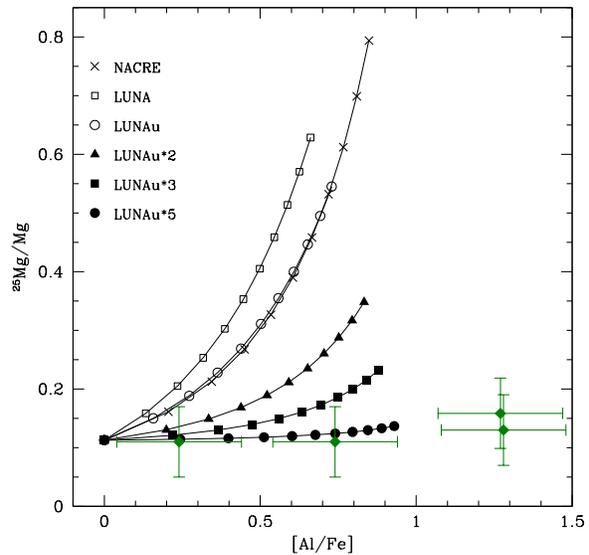}}
\vskip-60pt
\caption{The fraction of $^{25}$Mg with respect to the overall magnesium in the
gas ejected by a $5~M_{\odot}$ AGB model, run with the chemical composition of
M13 stars. The various symbols indicate the results obtained with different cross
sections of the proton capture reactions by $^{25}$Mg nuclei. Open circles, full
triangles, squares and points indicate, respectively, the results obtained by
considering the upper limits for the above reactions and the same upper limits
multiplied by a factor 2, 3 and 5. The various points along a given track indicate
the results obtained when mixing the pure AGB ejecta with variable percentages of
gas pristine, with $10\%$ steps. The green diamonds represent the results on M13 
giants by \citet{yong06}.}
\label{f25rela}
\end{figure}

The main features of the Mg-Al nucleosynthesis associated to HBB in massive AGB stars is
reported in Fig.~\ref{f5msun}, showing the evolution of the surface mass fractions of
the magnesium isotopes and of aluminium; the total magnesium is also shown. 
Fig.~\ref{f5msun} refers to a $5~M_{\odot}$ model calculated with the metallicity of
M13 ([Fe/H]=--1.5) and the same initial magnesium of FG stars in M13, 
[Mg/Fe]=+0.2.\footnote{This estimate is somewhat uncertain, given the large spread
in the magnesium abundances among the FG stars of M13 in the data by \citet{meszaros15}.
Other works on the same cluster find a higher initial Mg, suggesting a possible offset in 
the data. The choice of the initial 
Mg does not affect the overall Mg depletion caused by HBB, because the rate of Mg burning
scales linearly with the Mg mass fraction. The same holds for the relative fractions of
the various isotopes, provided that the same initial ratios are used. On the other hand,
Al production is influenced by the assumed initial Mg, because a higher availability of
magnesium leads to higher Al production. Therefore, in the results shown in
Fig. \ref{f25rela} and \ref{fluna}, the theoretical Al abundances must be
considered as lower limits.},
according to the recent observations by \citet{meszaros15}; 
within the framework of the self-enrichment mechanism
by AGB stars, it is a typical mass expected to provide the gas from which SG stars
formed. Notice that in the cluster NGC\,2419, the presence of SG stars
largely enhanced in helium suggests pollution from more massive AGB and/or super-AGB stars 
(6--8\Msun, Ventura et al. 2013). The HBB temperatures in this case are
T$ \sim 100-150$\,MK\footnote{The temperature of the
base of the envelope changes during the AGB phase: initially $T$ increases as the
core mass grows, whereas towards the end of the evolution $T$ diminishes, because of the
gradual loss of the external mantle. However, during the time when most of mass loss
occurs the temperature at the base of the external mantle of the stars is approximately 
constant, which allows us to define a typical HBB temperature.}.

We see in Fig.~\ref{f5msun} that $^{24}$Mg burning starts since the initial AGB
phases, leading to a significant drop in the $^{24}$Mg content and to the 
formation of $^{25}$Mg; the latter isotope reaches a maximum
and then declines. $^{26}$Mg is only marginally
touched by this nuclear activity, whereas $^{27}$Al is produced. Despite the depletion 
of the overall magnesium is within a factor $\sim 2$, this results in a significant
increase in the aluminium content, because magnesium is much more abundant than
aluminium.

A clear result from Fig.~\ref{f5msun} is that the ejecta of these stars will be greatly
enriched in $^{25}$Mg; it is indeed the accumulation of the latter isotope in the 
surface regions which prevents a higher depletion in the overall magnesium. The
central role played by $^{25}$Mg in the overall Mg-Al nucleosynthesis was underlined
by \citet{ventura11a}. The large $^{25}$Mg content expected in the ejecta of massive 
AGB stars was used by \citet{denissenkov15} as argument pointing against the 
self-enrichment by AGBs scenario.

\subsection{The status of the art in modelling Mg-Al-Si in AGB envelopes}

The comparisons between observations and model results are based on the assumption 
that the formation of SG stars occurs in gas formed by mixing of the 
polluters ejecta (in the present case, the AGB ejecta) with gas having the same composition 
of FG stars (pristine gas). The ratio of polluter to pristine gas may 
vary from 1 (pure ejecta) to zero (pure first generation gas). This hypothesis is 
necessary in all formation scenarios \citep{decressin2007b, denissenkov15}, as it is 
demanded by the shape of the correlations - anticorrelations patterns \citep[e.g.][]{carretta12a}. 
For the AGB scenario, the model has been applied to several clusters \citep{dercole2010, 
dercole2012, ventura2014}, finding a reasonable, although certainly not perfect, 
correspondence between the yields and the measured abundances \citep{dantona16}.

The strong nucleosynthesis activated via HBB at the bottom of the external mantle of
massive AGB stars, with the extreme sensitivity of the strength of HBB to the 
metallicity, allowed \citet{ventura16} to interpret the Mg-Al trends of GC stars
based on Apogee data, published in \citet{meszaros15}, as due to self-enrichment
by AGB stars belonging to the FG of the same clusters. The models presented in 
\citet{ventura16} could successfully reproduce the large Mg and Al spreads showed by 
M92 stars, the smaller Mg spread observed in M3 and the Mg-Al trends observed
in two more metal-rich clusters, M5 (no Mg-spread, significant Al spread) and
M107 (no spread in Mg and Al). The comparison with M13 data showed that the stars
with the most extreme chemical composition have Mg abundances $\sim 0.15-0.2$\,dex 
smaller than predicted by the AGB models of the appropriate metallicity.
The approach followed by \citet{ventura16} was extended to other Galactic GCs in the
recent study by \citet{flavia18}.

On the wake of the results by \citet{ventura16} and \citet{flavia18},
we may explain the different
depletion of magnesium and aluminium reported in Table \ref{tabmg} as due to the
gas ejected by massive AGB stars, which shows a higher degree of alteration with 
respect to the original chemistry for lower metallicity.
For GCs sharing the same metallicity, the most extended Mg-Al trends are detected 
in the clusters where some SG stars formed directly from genuine gas expelled by AGB stars, 
consistent with the large helium spread detected. 

We now make a step forward towards a full and satisfactory interpretation of
the chemical patterns observed in GCs, by attempting to reproduce qualitatively and
quantitatively the extension of the Mg-Al anti-correlation in the GCs harbouring 
stars with an extreme chemistry, formed from the gas expelled from polluter stars.
This analysis is more robust and complete in comparison with previous works on this
argument, because we test our findings against the relative distribution of the
magnesium isotopes, which is extremely sensitive to the temperature at which 
the gas processed by p-capture nucleosynthesis, from which SG stars formed, was
exposed.

Apart from supermassive stars (see Section \ref{disc}), the Hot Bottom Burning envelopes of AGBs 
are the only stellar p-burning environment  reaching the high temperatures necessary to 
process $^{24}$Mg \citep{prantzos07}, but today's p--capture rates provide 
ratios of the Mg isotopes which are at variance with the available observational data.
It is therefore necessary to work out whether ratios consistent with the available data 
can be obtained by varying the relevant cross sections. In the end, we may ask 
whether the required variations are compatible with the uncertainties of the 
cross sections determinations, and new measurements may be planned to check these 
predictions.  

Based on the arguments given above, we will primarily focus our analysis on M13 and NGC 6752,
two clusters for which isotopic ratios in stars with significantly different
total magnesium abundances are available. We will also consider NGC 2419, as it is the 
cluster where the largest Mg spread has been detected so far. 
We are not discussing the results on M4 stars by \citet{dacosta13}, because
the two stars observed show a very large overall $[Mg/Fe]=+0.4$ and very similar isotopic 
ratios, thus suggesting that they both belong to the FG of the cluster.

\section{Model computation}
\label{results}
The results regarding the Mg-Al nucleosynthesis used in the previous papers
by our group, including \citet{ventura16} and \citet{flavia18}, 
have been based on the NACRE \citep{angulo99}
cross sections for proton capture reactions by magnesium and aluminium isotopes. To 
infer which is the largest depletion achievable for a given metallicity, we used the 
NACRE upper limits for the reaction rates of proton captures by $^{25}$Mg, $^{26}$Mg and 
$^{26}$Al. 

The NACRE rates have been revised and we are interested into a more detailed analysis, 
aimed at understanding the variation of the individual isotopes, besides the extent of the 
overall depletion of magnesium. Therefore, we fully update the rates, by adopting the 
STARLIB library \citep{sallaska13}, completed by the newest results from the LUNA 
collaboration for what attains the cross sections of the proton capture reactions by 
$^{25}$Mg nuclei \citep{straniero13}.

\subsection{M13}
To study the Mg-Al nucleosynthesis in M13 we calculated several evolutionary
sequences of the $5~M_{\odot}$ model shown in Fig.~\ref{f5msun}, with various
assumptions regarding the cross sections of the proton capture reactions
by the isotopes involved in the Mg-Al chain. The chemical composition is the same
as given in \citet{meszaros15} for FG stars in M13, i.e. [Fe/H]=--1.5, [O/Fe]=+0.55, 
[Mg/Fe]=+0.2 and [Si/Fe]=+0.4. The mass fractions of all the other species are assumed 
to be solar scaled. These choices lead to the metallicity Z=10$^{-3}$.

In Fig.~\ref{f25rela} we show the expected variation of the $^{25}$Mg/Mg ratio versus Al 
in the gas ejected; the quantities shown in the plane reflect the average chemical
composition of the gas expelled by the star during the whole AGB phase. In the figure 
we show a dilution curve, obtained by assuming various degrees of 
mixing between the material lost by the star via stellar winds and pristine matter, having the
same chemical composition as the FG stars in the same cluster. The observations
by \citet{yong06} are also reported in the same plane.

The results in Fig.~\ref{f25rela} confirm that when using the NACRE
reaction rates, the Mg content of the ejecta is dominated by $^{25}$Mg
($\sim 80\%$ of the total Mg). To make the fraction of $^{25}$Mg compatible
with the observations, i.e. $\sim 10-30\%$, we must consider a dilution
with at least $70\%$ pristine gas; this is even less compatible with the
observational evidence, as the Al-enhancement would be $\delta [Al/Fe] < 0.5$.
Fig.~\ref{f25rela} also shows that adopting the recommended values, or the upper 
limits of the LUNA cross sections for the two proton capture reactions by $^{25}$Mg, 
does not improve the agreement with the data. 

We also plot the results obtained by
artificially multiplying the LUNA cross sections by different factors.
The slope of the various curves is flatter the higher are the rates adopted,
owing to the lower equilibrium abundances of $^{25}$Mg obtained for higher rates. 
The results shown in Fig.~\ref{f25rela} indicate that, as far as the relative
fraction of $^{25}$Mg is concerned, the agreement with the results by
\citet{yong06} requires that the LUNA cross sections must be increased by 
at least a factor 3. We note that while the results shown in Fig.~\ref{f25rela} 
have been obtained 
by enhancing by the same factor both the $^{25}$Mg burning channels, we would come to 
similar conclusions if we had modified solely the slower reaction giving the time
scale of $^{25}$Mg burning, i.e. the $^{25}{\rm Mg}(p,\gamma)^{26}{\rm Al^{m}}$ reaction.
This follows the discussion in section 3.1.

\begin{figure*}
\begin{minipage}{0.48\textwidth}
\resizebox{1.\hsize}{!}{\includegraphics{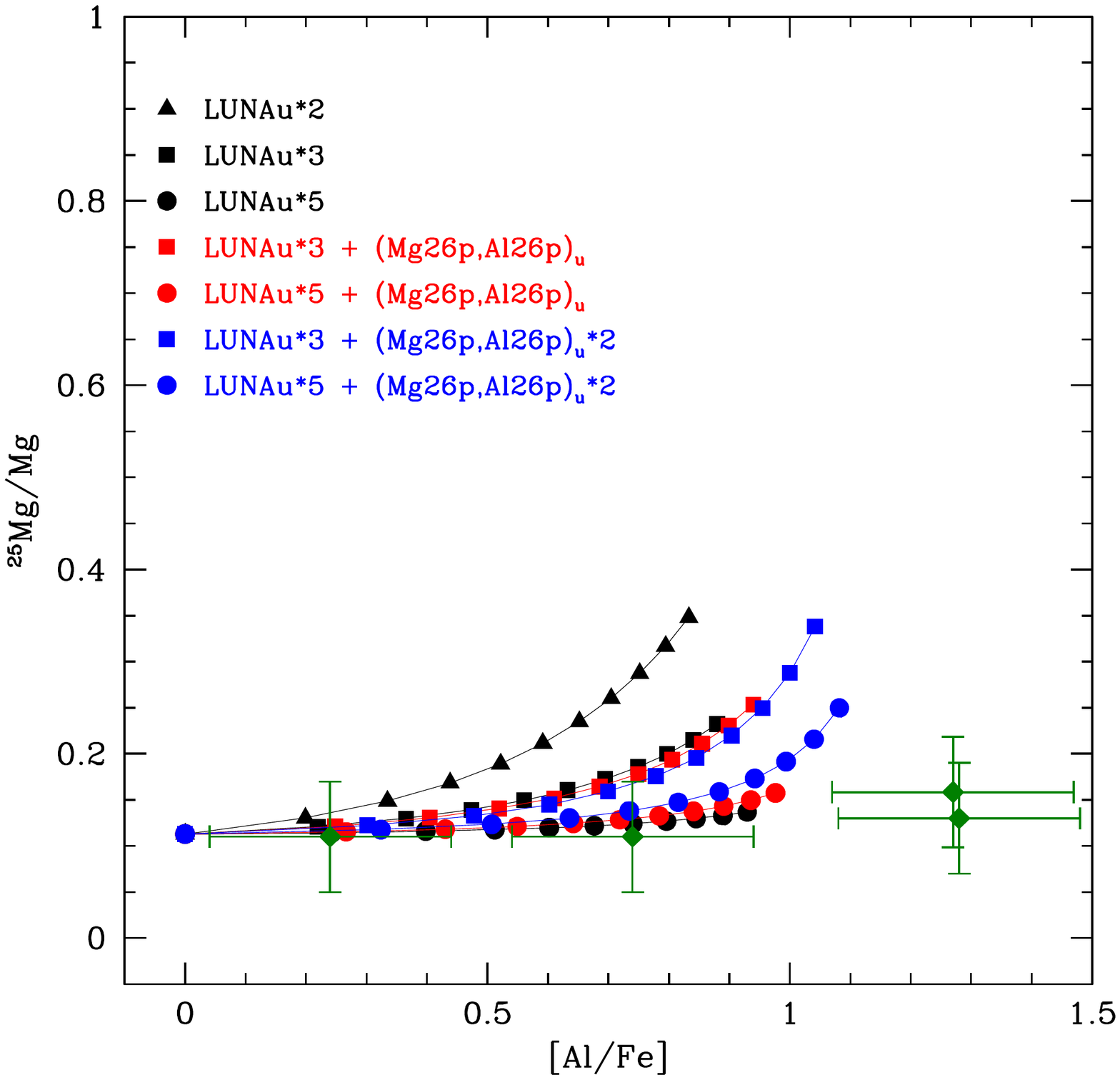}}
\end{minipage}
\begin{minipage}{0.48\textwidth}
\resizebox{1.\hsize}{!}{\includegraphics{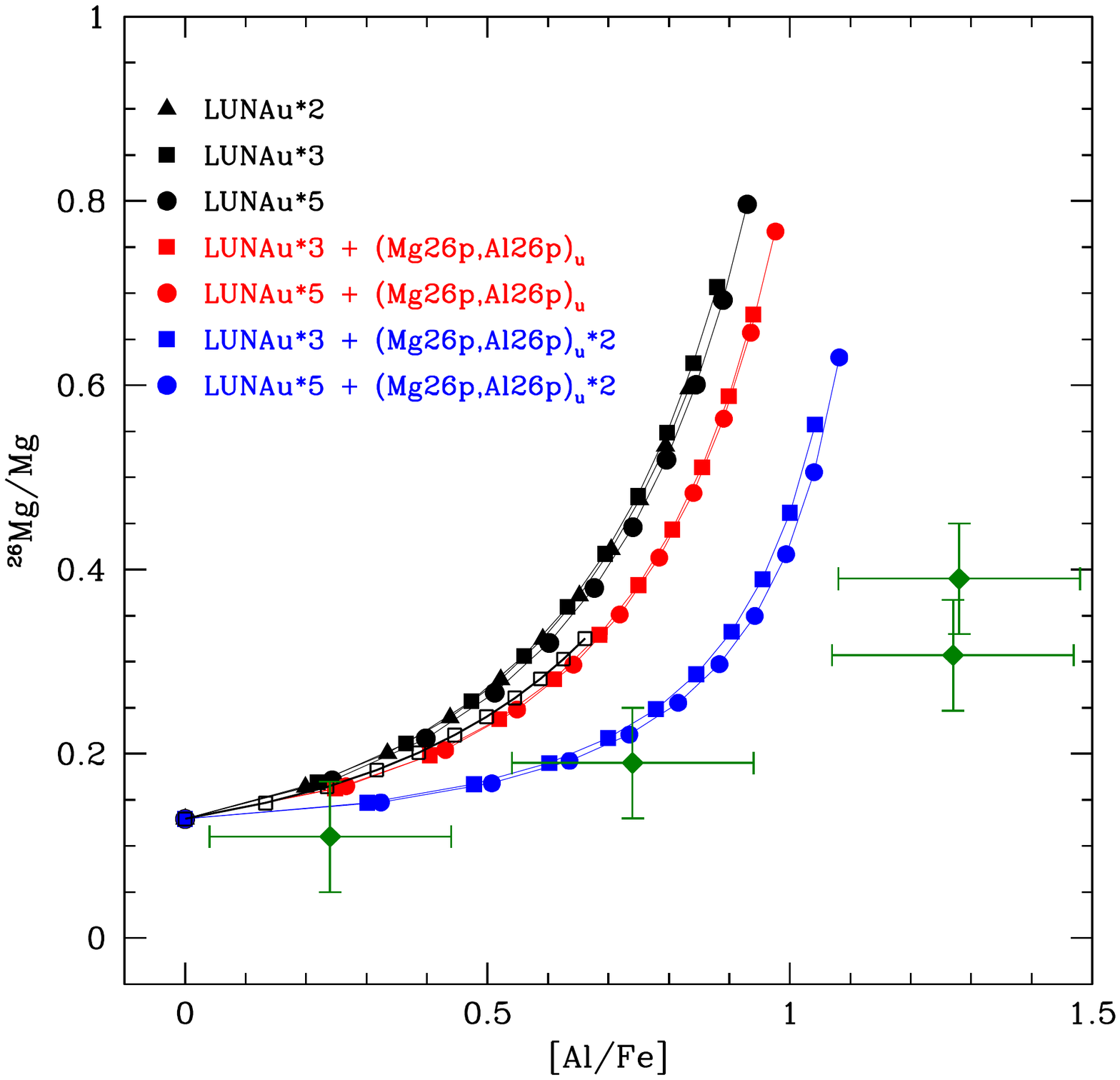}}
\end{minipage}
\vskip-80pt
\begin{minipage}{0.48\textwidth}
\resizebox{1.\hsize}{!}{\includegraphics{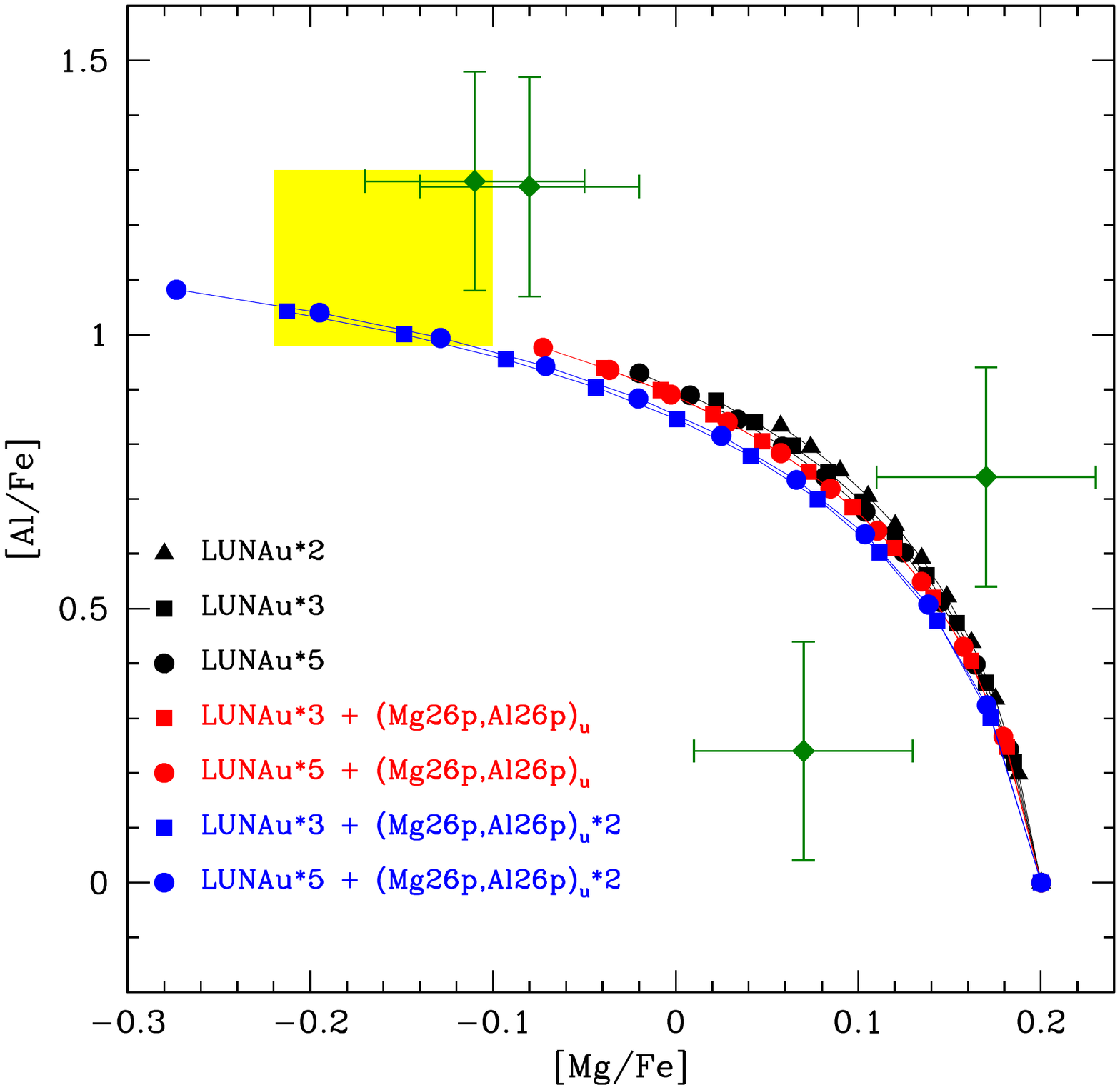}}
\end{minipage}
\begin{minipage}{0.48\textwidth}
\resizebox{1.\hsize}{!}{\includegraphics{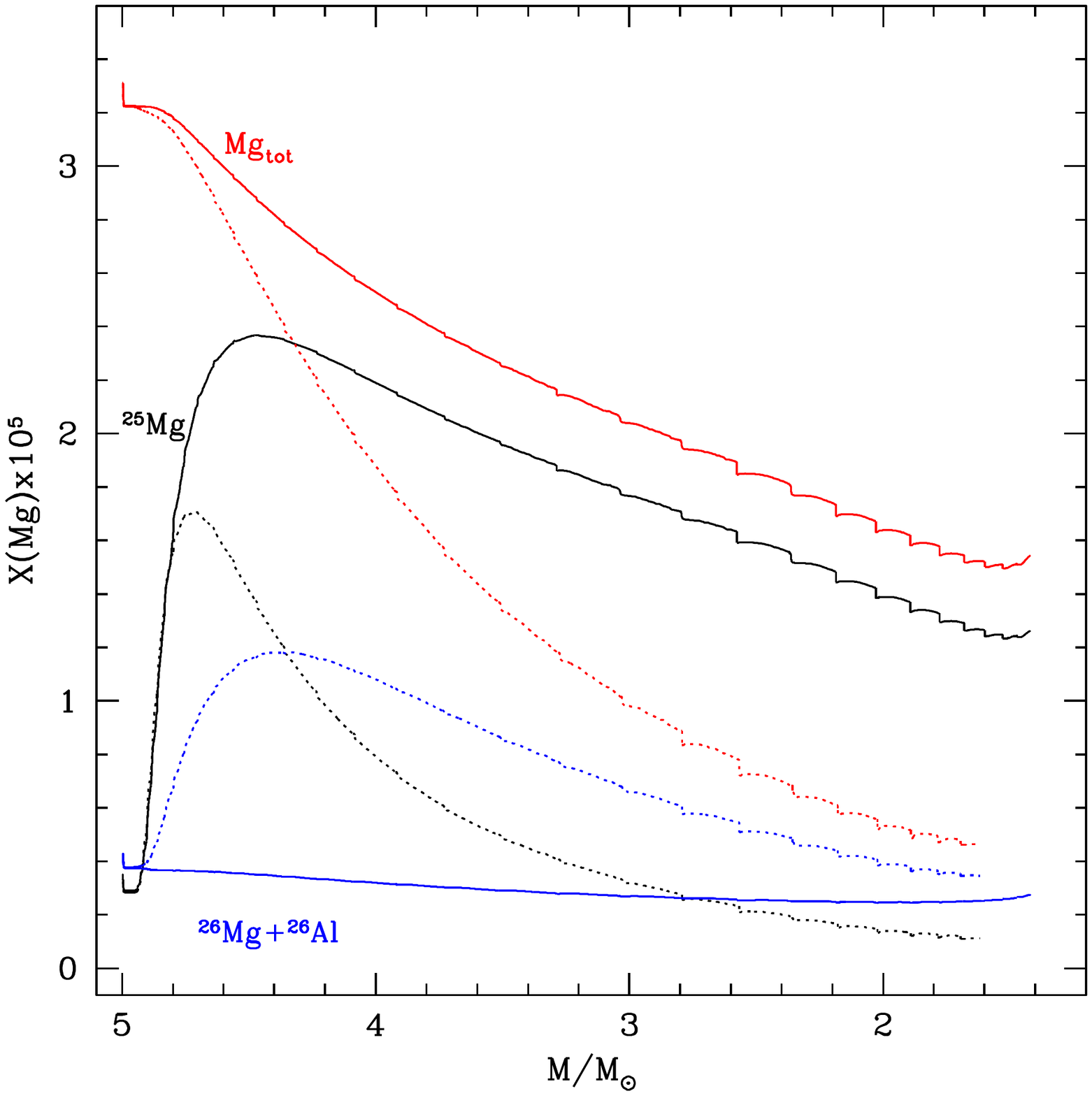}}
\end{minipage}
\vskip-50pt
\caption{The relative fractions of $^{25}$Mg (top, left panels) and $^{26}$Mg
(top, right) and of the Mg-Al mass fractions (bottom, left) in the gas lost by the 
$5~M_{\odot}$ model shown in Fig.~\ref{f5msun}, for various choices of the relevant 
cross-sections. We show the dilution curves, obtained by mixing the AGB gas with various 
fractions (from 0 to $100\%$, with $10\%$ steps) of pristine matter. The results on M13
giants by \citet{yong06} are indicated with green diamonds. The yellow shaded region
point the zone in the various planes populated by M13 stars with the most extreme
chemical composition by \citet{meszaros15}. In the bottom, right
panel we report the sensitivity to the choice of the cross-sections of the evolution of 
the magnesium isotopes and of aluminium, for the same model described in
Fig.~\ref{f5msun}: the solid tracks correspond to the results shown in Fig.~\ref{f5msun},
whereas dotted lines indicate the results obtained when adopting the same cross sections
as the full, blue squares in the other panels. 
}
\label{fluna}
\end{figure*}

Fig.~\ref{fluna} shows the relative fractions of $^{25}$Mg and $^{26}$Mg and the overall
Mg-Al trend, for the same $5~M_{\odot}$ model discussed in the previous figures. 
As in Fig.~\ref{f25rela}, we focus on various combinations of cross-sections, with the 
LUNA rates for $^{25}$Mg burning multiplied by a factor of 2 or more.

The results in Fig.~\ref{fluna} indicate that an ad hoc increase in
the LUNA rates are not sufficient to achieve agreement between models and 
observations, because adopting higher rates for proton captures by $^{25}$Mg
nuclei, while decreasing the relative fraction of $^{25}$Mg, would favour
extremely large $^{26}$Mg$/$Mg$> 60\%$, at odds with the values found by 
\citet{yong06}. This can be seen in the right, top panel of Fig.~\ref{fluna},
where we note the large values attained by $^{26}$Mg$/$Mg when using the
enhanced LUNA rates with the recommended STARLIB rates for proton captures by
$^{26}$Mg and $^{26}$Al (black dots).

This problem is somewhat alleviated when the upper limits for the same
reactions are adopted (red dots in the same panel), whereas the
agreement is satisfactory when the cross sections of the reactions 
involving $^{26}$Mg and $^{26}$Al are multiplied by a factor of 2
(blue dots in Fig.~\ref{fluna}).

This choice also allows fixing the problem of the overall depletion of 
magnesium detected in M13 stars with the most extreme chemistry, discussed 
in \citet{ventura16}: this can be seen in the left, bottom panel of Fig.~\ref{fluna},
showing that a total depletion $\delta($Mg$)\sim -0.4$ is obtained when using 
the combination of the LUNA cross sections increased by a factor $\sim 3-5$ 
and the STARLIB rates for proton captures by $^{26}$Mg and $^{26}$Al
enhanced by a factor 2.

In the right, bottom panel of Fig.~\ref{fluna} we compare the results shown in
Fig.~\ref{f5msun} with those obtained with the combinations of LUNA and
STARLIB rates discussed above. We note the significant differences in the
behaviour of $^{25}$Mg and, as a consequence, in the overall magnesium.

\begin{figure*}
\begin{minipage}{0.32\textwidth}
\resizebox{1.\hsize}{!}{\includegraphics{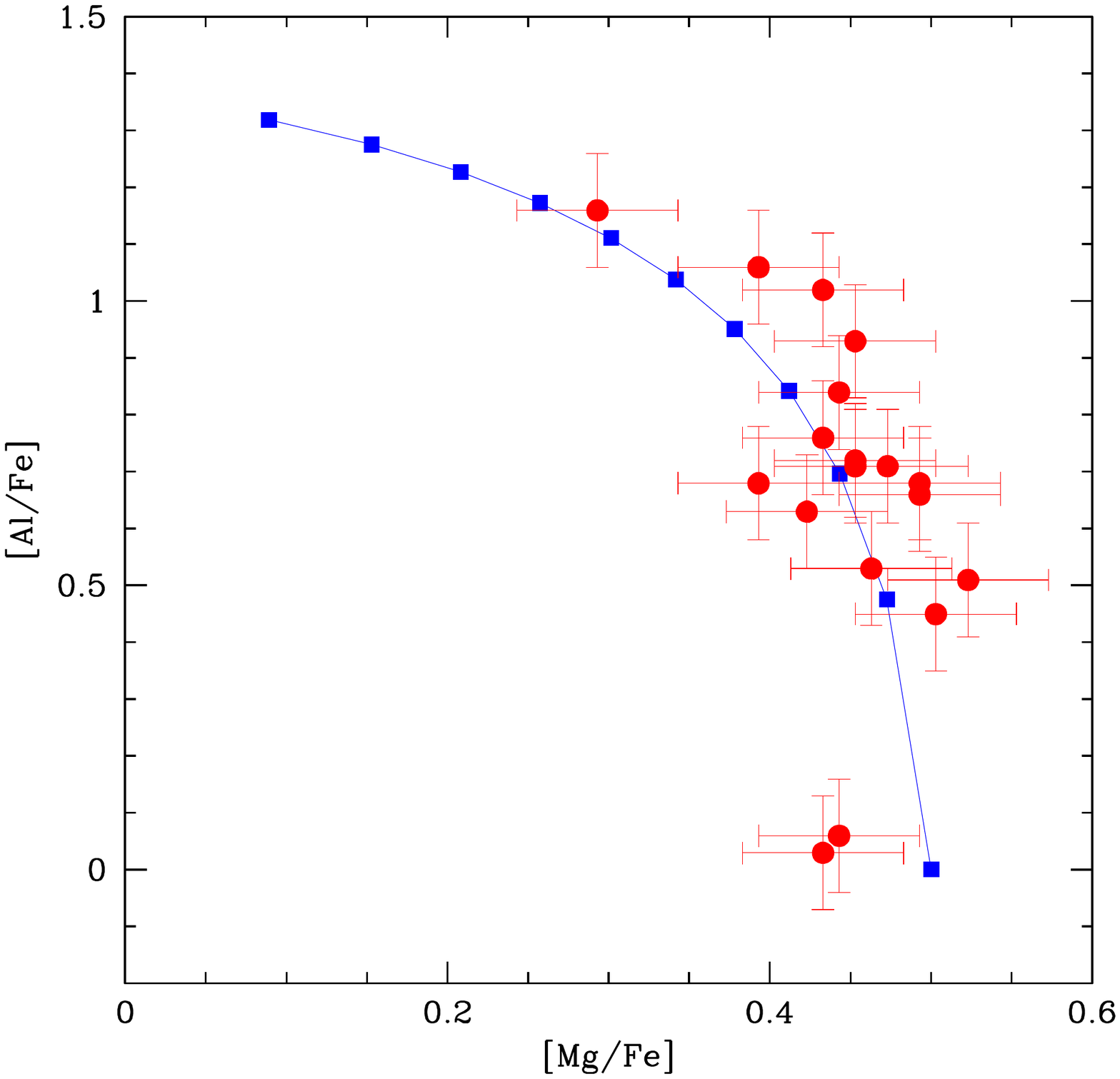}}
\end{minipage}
\begin{minipage}{0.32\textwidth}
\resizebox{1.\hsize}{!}{\includegraphics{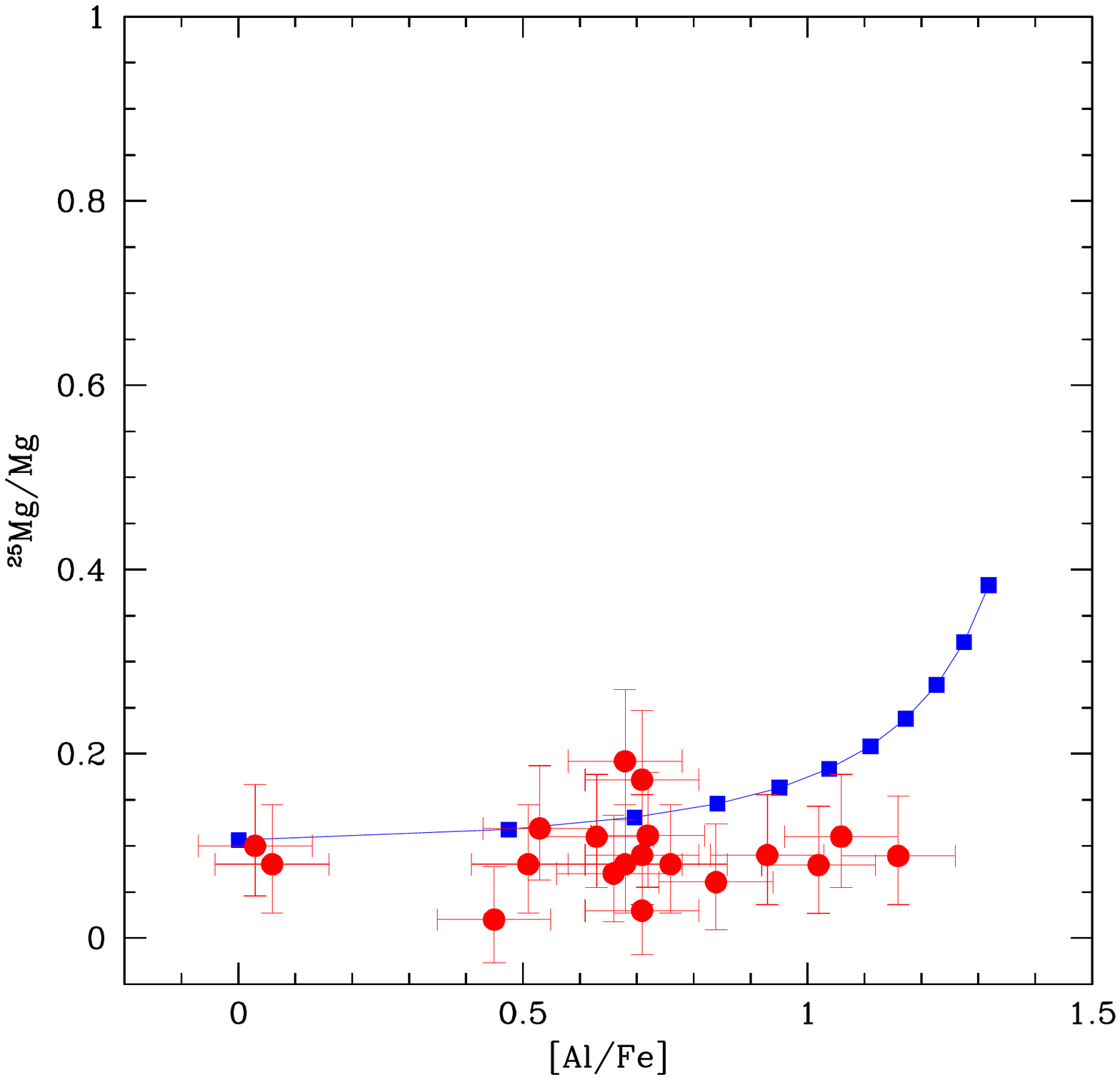}}
\end{minipage}
\begin{minipage}{0.32\textwidth}
\resizebox{1.\hsize}{!}{\includegraphics{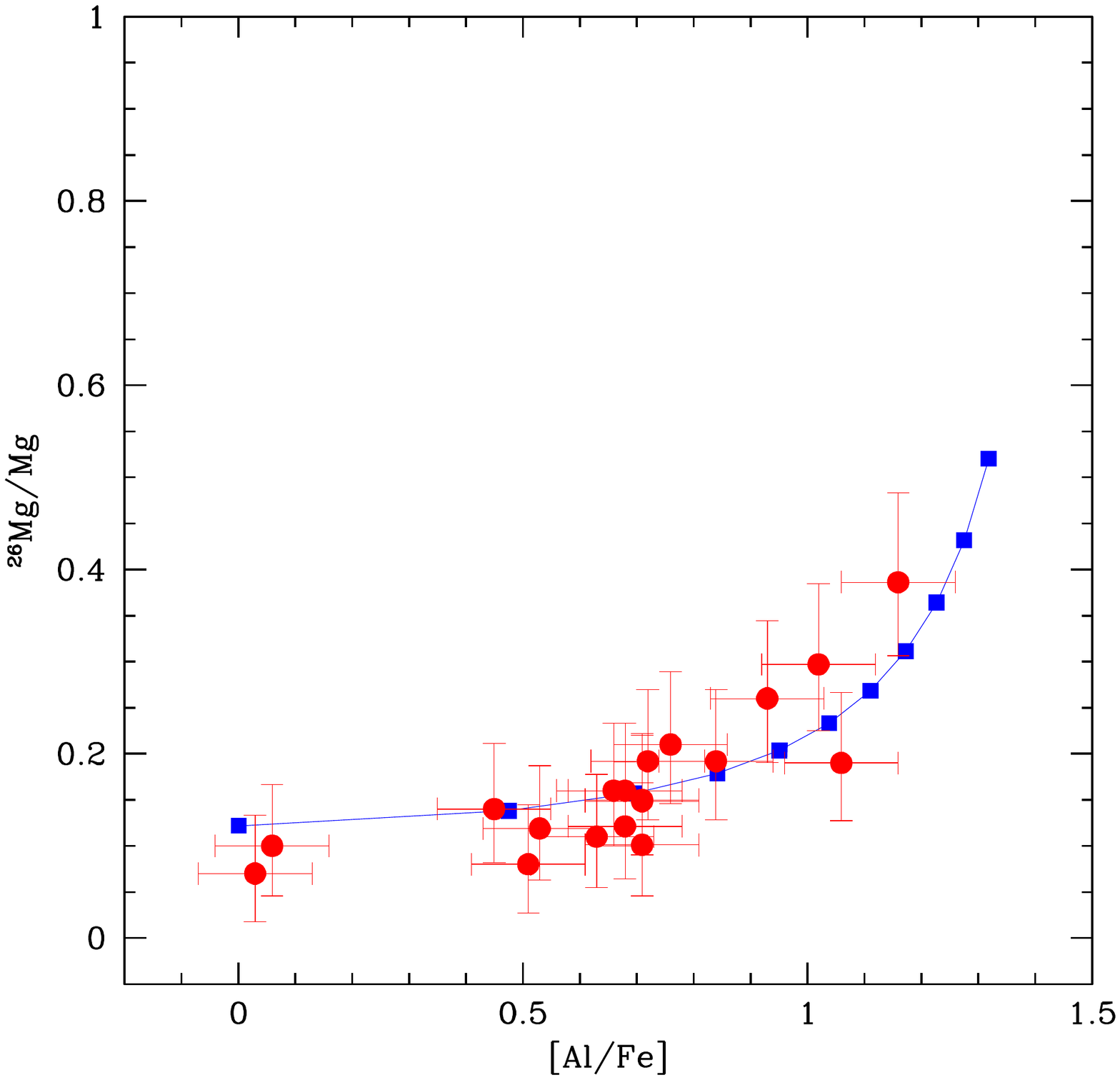}}
\end{minipage}
\vskip-30pt
\caption{The observation of NGC 6752 by \citet{yong03} overimposed to results from
AGB modelling regarding the Mg-Al abundances (left panel) and the relative fractions of
$^{25}$Mg (middle) and $^{26}$Mg (right). As in the previous figures we show the 
results obtained by diluting the AGB material with pristine gas.
}
\label{6752}
\end{figure*}

\subsection{NGC 6752}
We tested the combination of cross sections used for M13 in the previous section
against the results on the Mg and Al abundances of NGC 6752 stars by \citet{yong03};
to this aim, we calculated AGB models with the same metallicity as M13 ([Fe/H]=--1.5), but with a
higher initial magnesium ($[Mg/Fe]=+0.5$) than M13. 

We compare the chemical composition of the ejecta from AGB stars with the observations 
in Fig.~\ref{6752}; the three panels of the figure show the dilution curve of pure AGB gas
mixed with different fractions of pristine gas, with the same chemistry of FG stars.

On the Mg-Al plane, shown in the left panel of Fig.~\ref{6752}, the observations are well 
reproduced by the models. These results indicate that SG stars in this cluster formed
from AGB gas mixed with pristine gas. The chemical composition of the star with the most 
extreme chemistry, i.e. with the largest Al and the smallest Mg, is compatible with
AGB gas diluted with $\sim 40\%$ of pristine gas.

The comparison of the models with the relative fractions of $^{25}$Mg and $^{26}$Mg
is also pretty satisfactory. The dilution curves in the middle and right panels
of Fig.~\ref{6752} reproduce the observed trends, with the same degree of mixing
required to fit the Mg-Al trend. 

The analysis of NGC 6752 stars outlines a clear difference compared to M13, because in 
this case a significant fraction of pristine gas (more than $\sim 30$\%) is required.
It is possible that the sample we are examining \citep{yong03} does not include stars with 
more extreme chemistry \citep{grundahl2002}, and in fact the horizontal branch morphology 
of the cluster stars requires a variation (SG-FG) of helium up to $\delta$Y$\sim$0.07 (Tailo et 
al., in preparation), while the ``average" width of the MS is only $\delta$Y$\sim$0.04 
\citep{milone13}.

In the models which deplete Mg, the nucleosynthesis proceeds up to silicon. The mild 
correlation between Al and Si abundances found by \cite{yong03} can be compared with that 
predicted by these models. The observed  $\delta$[Si/Fe] between the abundance in models 
with standard Al and the average abundance in models with high Al is $\sim 0.04$\,dex. 
In the ejecta of the 6\Msun\ we find $\delta$[Si/Fe]$\sim$+0.035\,dex, so for a dilution 
of 40\% the total variation expected is $\sim$+0.02. We can regard this comparison as a 
reasonable agreement, in view of the large uncertainties in Silicon production and 
Magnesium depletion discussed in \cite{ventura11a}.

\begin{figure*}
\begin{minipage}{0.48\textwidth}
\resizebox{1.\hsize}{!}{\includegraphics{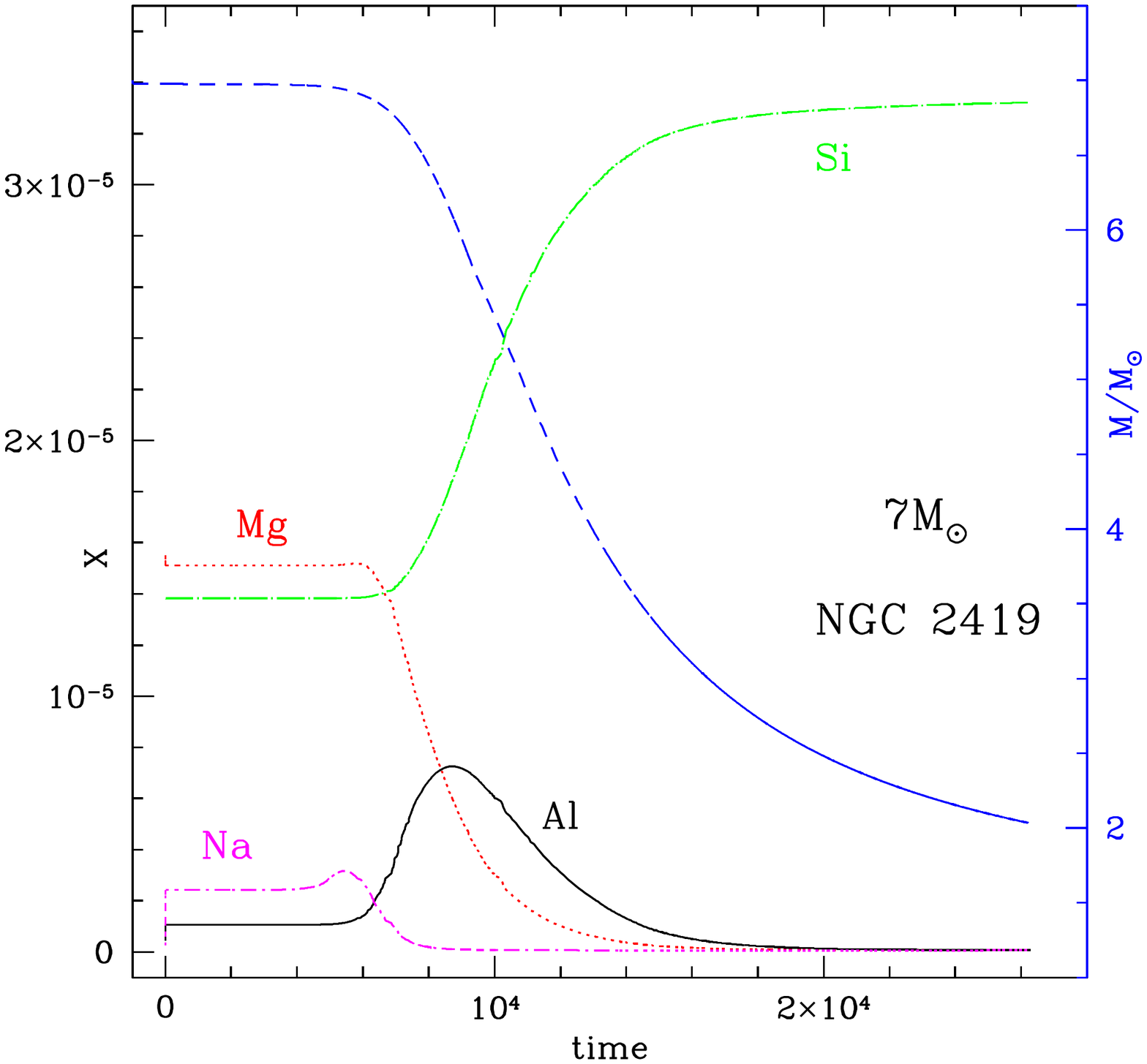}}
\end{minipage}
\begin{minipage}{0.48\textwidth}
\resizebox{1.\hsize}{!}{\includegraphics{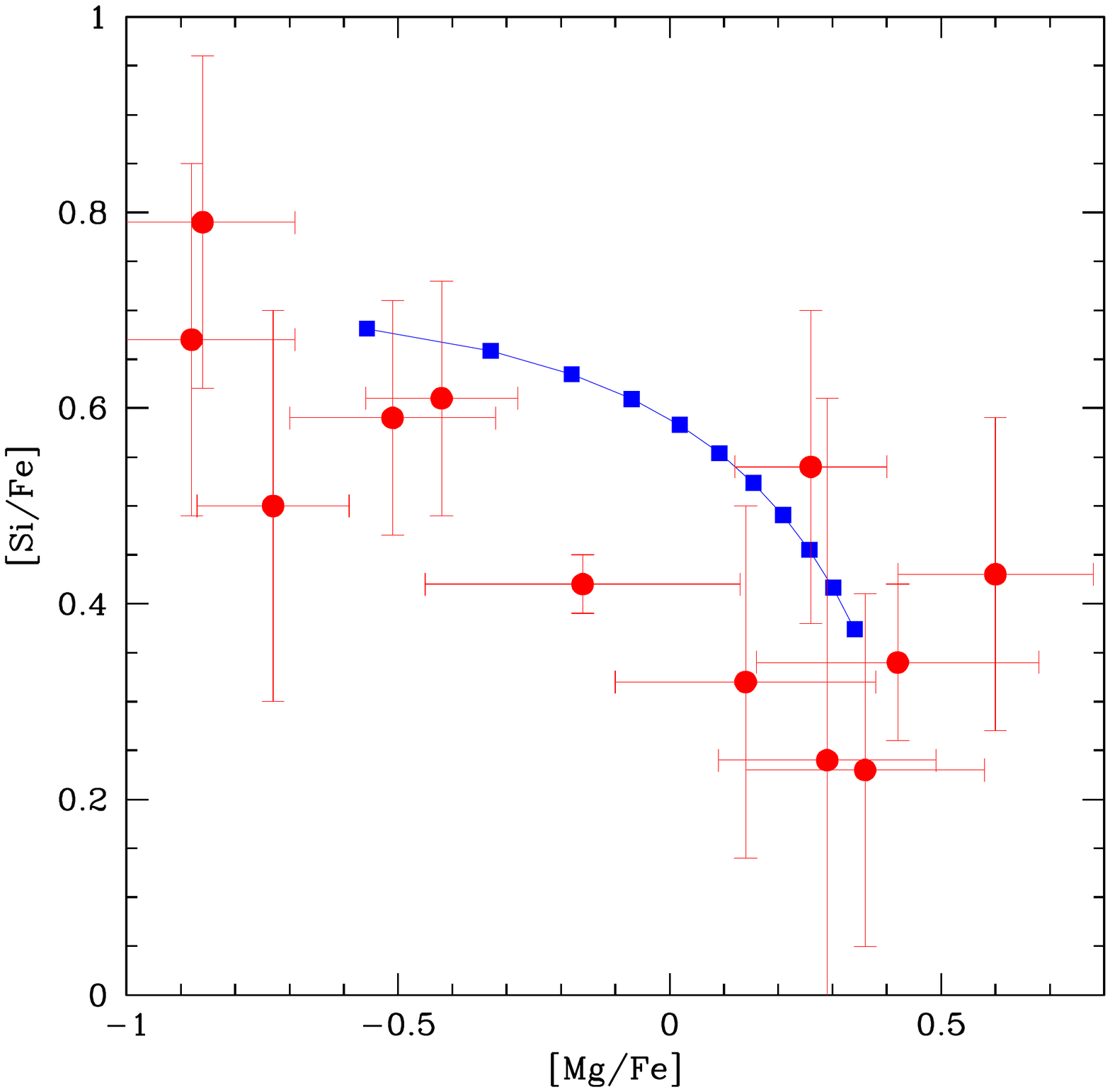}}
\end{minipage}
\vskip-60pt
\caption{Left: The AGB evolution of the total magnesium (red, dotted line), aluminium
(black, solid), silicon (green, dotted-dashed) and sodium (long-dashed, magenta) in
a $7~M_{\odot}$ model with the same chemical composition of stars in NGC 2419.
Right: The dilution pattern of the Mg-Al content in the AGB models presented in the
left panel, overimposed to the observations by \citet{cohen12}.
}
\label{f2419}
\end{figure*}

\subsection{NGC 2419}
NGC 2419 is a very peculiar metal-poor GC ($[Fe/H] = -2.1$, see Table \ref{tabmg}), 
as witnessed by the complex morphology of the HB and the chemical patterns traced by 
the distribution of the observed mass fractions of some species, determined by 
high-resolution spectroscopy.

The analysis by \citet{marcella15} showed that the HB of NGC 2419 can be explained 
only by invoking the presence of a helium-rich population, with $Y > 0.35$. 
In the AGB scenario, this is consistent with the
formation of SG stars directly from the winds of very massive AGB stars, the progeny of
stars with initial mass above $\sim 6~M_{\odot}$. Models for the formation of an extreme 
SG from undiluted AGB ejecta have been proposed by \citet{dercole08, dercole16} and
\citet{dantona16}. Dilution with pristine gas, if any, must have been negligible in this 
case, otherwise the helium content of these SG stars would be smaller, as a consequence 
of mixing of helium-rich matter from the AGB winds with $Y \sim 0.25$ pristine gas.

Furthermore, the results from spectroscopy outlined the presence of an extremely large
Mg-spread \citep{cohen12} and of a clear magnesium-potassium anticorrelation
\citep{mucciarelli12}. An interesting point on this side is that the distribution of
the stars in the Mg-K plane is bimodal, with a group of stars with large Mg and
solar scaled K, likely the FG of the cluster, well separated by SG stars, which
exhibit a large Mg depletion ($\delta [Mg/Fe] \sim -1$ dex compared to FG stars) and
K enhancement ($\delta [K/Fe] \sim +1$ dex). \citet{ventura12} suggested that
such an extreme chemical composition could be achieved in the interior of massive,
metal-poor AGB stars, owing to the effects of a very strong HBB.

The independent analysis by \citet{iliadis16}, aimed at fixing the thermodynamic
conditions compatible with the nucleosynthesis required to reproduce the most extreme
chemistries of NGC 2419 stars, confirmed that the temperatures and densities achieved at 
the base of massive AGB models of the same metallicity of NGC 2419 can account for the
chemical composition of the stars with the most extreme chemistry.

In spite of the lack of Mg isotopic ratios determinations, it is interesting to test the 
AGB models presented here against the Mg spread observed in  NGC 2419, because an 
evaluation of the reliability of the theoretical description given here can be obtained by 
comparing the chemical composition of the ejecta of massive AGB stars with the chemistry 
of SG stars in the cluster.  As there has been very little or no dilution with pristine 
gas, a direct comparison between models and observations is possible.

To this aim, we calculated massive AGB models with the metallicity of NGC 2419 stars, 
with the O, Mg and Si observed in FG stars of this cluster: $[O/Fe]=+0.4$, $[Mg/Fe]=+0.4$, 
$[Si/Fe]=+0.35$. This mixture and the assumed $[Fe/H]$ correspond to the metallicity
$Z=2\times 10^{-4}$. The left panel of
Fig.~\ref{f2419} shows the temporal variation of the surface chemical composition of
a $7~M_{\odot}$ star, during the AGB phase. Owing to the effects of the second dredge-up,
the helium content of the gas ejected by this star is $Y = 0.37$, in agreement with the
analysis by \citet{marcella15}. We focus on the surface abundance of
Na, Mg, Al, Si, the only elements involved in p-capture nucleosynthesis for which we
have data from \citet{cohen12}. In the same figure we also show the 
evolution of the mass of the star (scale on the right), to have an idea of the chemical 
composition of the ejecta. 

The results shown in the left panel of Fig.~\ref{f2419}, can be summarized as follows:

\begin{enumerate}
\item{The overall magnesium is destroyed at base of the envelope of massive AGB stars
with the metallicity of NGC 2419, via strong HBB. The destruction of magnesium
begins after $\sim 20\%$ of the AGB phase has past; because during this time little
mass was lost by the star, the gas ejected is magnesium-poor.
}
\item{When Mg burning starts, Al-production occurs. However, owing to the very hot
HBB temperatures, in later phases Al is destroyed by proton fusion, in favour of
silicon. We expect only a moderate increase in Al in the gas expelled.
}
\item{The combination of Mg and Al burning leads to the synthesis of silicon. The final
silicon is almost a factor 3 higher than the silicon initially present in the star.
}
\item{The HBB temperatures at which the base of the envelope is exposed are too large to
allow the formation of great quantities of sodium. The initial increase in the surface
sodium triggered by the second dredge-up and by proton capture by $^{22}$Ne nuclei is
followed by a phase during which sodium is destroyed owing to proton capture. The average
sodium in the gas ejected is only slightly increased with respect to the initial
chemical composition.}
\end{enumerate}

In the right panel of Fig.~\ref{f2419} we show the Mg--Si observations by \citet{cohen12}
compared to the chemistry of the ejecta; the results for various degrees of dilution
of the pure AGB gas with pristine matter are indicated. We choose the Mg-Si plane
to make this comparison, because the Al is available only for part of the stars in
the \citet{cohen12} sample and no error bars are given for this element.
 
The overall Mg spread of the $7~M_{\odot}$ model reproduces the difference between
the average Mg of FG and SG stars in NGC 2419; this is also in agreement with the
main finding by \citet{mucciarelli12}, who found a factor $\sim 10$ difference in the
Mg of FG and SG stars. This is a further confirmation that the combination of cross sections
required to fit M13 data allow to reproduce the large Mg spread shown by NGC 2419 stars.

Regarding the chemical composition of SG stars in this cluster, we note that the
small Al and Na spread between FG and SG stars observed by \citet{cohen12}, as also
the $\delta [Si/Fe] \sim 0.3$ silicon spread, are consistent with the results summarized
in the points (i)-(iv) given above.

\begin{figure*}
\resizebox{1.\hsize}{!}{\includegraphics{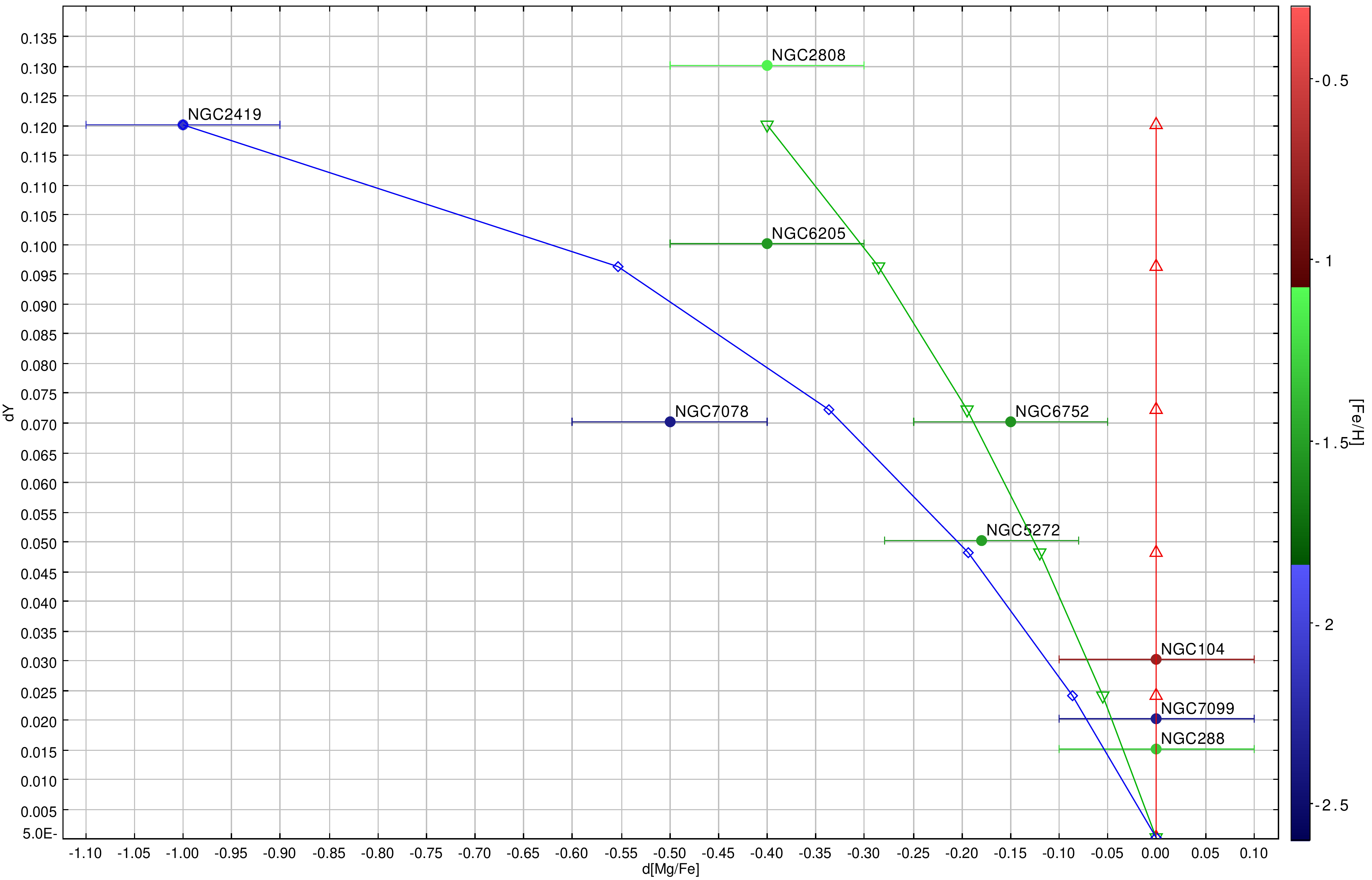}}
\vskip+10pt
\caption{The magnesium and helium spread for the clusters reported in Table \ref{tabmg},
for which the helium spread has been estimated. The colour coding corresponds to
different metallicities, as reported on the right, vertical axis. The dilution curves
were obtained by mixing the AGB ejecta with various percentages of pristine gas for
the metallicities $[Fe/H]=-0.77$ (red), $[Fe/H]=-1.12$ (green) and $[Fe/H]=-2.1$ (blue).
}
\label{felio}
\end{figure*}

\section{Global comparison with data}
\subsection{The AGB scenario, with the revision proposed for reaction rates}
Fig.\,\ref{felio} shows the data of Table\,\ref{tabmg}  compared with the yields of HBB 
in the above computations, in which the reaction rate of the proton capture reactions 
on $^{25}$Mg has been multiplied by three and those on  $^{26}$Mg and $^{26}$Al nuclei 
have been multiplied by a factor two. We plot the $\delta$Mg versus $\delta$Y data, 
assuming that the magnesium determination is affected by global errors of 0.1\,dex. 
We do not show the indermination in the Y values, but remember that these are derived by 
model interpretation of CM diagram features such as the HB and the MS width, so a fair 
error on the largest numbers would be at least $\delta$Y=$\pm$0.01, while in the clusters 
where small values are found, these values are compatible with zero. The three lines 
correspond to dilution curves for the metallicities of the computed models. 
With the limit of the modelling, the figure shows two main observational trends, which 
provide a strong  indication that indeed the AGB scenario describes correctly the chemical 
anomalies concerning magnesium.

\begin{enumerate}

\item The Mg and helium spreads are correlated: in the  clusters having a small He spread, 
e.g. M30 and 47Tuc, no Mg variation is detected.  The largest variations are found in 
clusters also harboring a very helium rich population. This reflects the role of dilution: 
a small helium spread suggests that no He-rich stars formed, i.e. that the contaminated gas, 
enriched in helium, was diluted with a significant fraction of `pristine' gas, whose 
composition is the same or close to the chemistry of the FG stars. 
In this interpretation, it is not surprising that no significant Mg spread is 
observed. If the processed Mg--poor gas lost by the polluters was mixed with Mg-rich,
pristine gas, the SG stars formed will have a very small (if any) magnesium depletion. 
Provided that strong dilution partly erases the effects of the nucleosynthesis, we understand that
the clusters we should look at to infer the most complete and exhaustive information
on the possible polluters are those exhibiting the most extended helium spread, because
they likely harbor (part of) SG stars formed without any dilution, whose chemical
composition reflects more directly the effects of the nucleosynthesis at which the
contaminating gas was exposed. Here we also remark that such a direct correlation with the 
helium abundance is not expected for sodium, because the sodium yield is not a result of 
pure destruction in HBB, but of a complex interplay between the 2DU of sodium, the early 
conversion of the $^{22}$Ne, acquired at the 2DU, into sodium by proton capture, and the 
sodium burning during HBB \citep{vd06}.

\item The data reported in Fig.~\ref{felio} show a clear trend with metallicity: 
among the clusters harboring stars greatly enriched in helium,
the Mg spread gets wider the lower is $[Fe/H]$. The cluster presenting the largest difference
between the Mg measured in FG and SG stars ($\delta [Mg/Fe] \sim -1$) is NGC 2419, a metal-poor 
cluster ($[Fe/H] = -2.1$) in which the presence of helium rich stars was proved beyond any 
reasonable doubt. NGC 2808 and M13, also believed to harbor a helium-rich population,
exhibit a narrower Mg spread compared to NGC 2419, i.e. $\delta [Mg/Fe] \sim -0.4$; these
clusters are more metal rich than NGC 2419, thus confirming the considerable sensitivity of
the Mg depletion to the iron content found in the modelling.
\end{enumerate}

In the context of the AGB scenario, the trends of the extension of Mg depletion 
with respect to both dilution and metallicity displayed in Fig.~\ref{felio} apply also to 
the less explored elements silicon and potassium. In particular, the recent results 
concerning the K dispersion \citep{mucciarelli17} are in perfect agreement with this 
scenario.

\subsection{The supermassive star models}

The data and this new modelling offer a valuable opportunity to discriminate among the
various pollutors that might have produced the gas from which SG stars formed in GCs. A 
successful model is required to produce  advanced Mg-Al nucleosynthesis during H--burning, 
so to produce Mg-poor, He--rich material, {\it in a modality strongly sensitive to 
metallicity}.

This evidence seems to indicate that massive binaries and fast rotating massive
stars could hardly have played a role in this context, because the core temperatures of these 
stars during core hydrogen burning, mainly determined by hydrostatic conditions requirements,
thus scarcely sensitive to the details of stellar modelling, are not sufficient to start 
Mg proton captures, and to produce gas with a Mg content comparable to the one observed 
in the most metal poor GCs, such as NGC 2419 (see the detailed discussion in section 2.2
in D'Antona et al. (2016)).

In section \ref{results} we showed that the yields of massive AGB models can account
for the observational evidence, being able to reproduce the Mg spread observed in
M13 and NGC 2419, and its trend with metallicity. However the observations of the magnesium 
isotopes, in the few clusters where these have been observed so far, are consistent 
with AGB yields only if the rates of the $^{25}$Mg and $^{26}$Mg p--captures are enhanced 
above the formal determination boundaries. 

On the contrary, processing in the cores of supermassive stars with masses of the order of 
$\sim 10^4~M_{\odot}$\ make a good job in reproducing the depletion and isotopic ratios 
{\it without any cross section adjustment}. Thus the supermassive stars scenario 
\citep{denissenkov14}, on these grounds, looks like it can be better than the AGB scenario.
\citet{denissenkov14} show that masses $\sim 10^4~M_{\odot}$\, reach core temperatures of the
order of $\sim 70-80$ MK, sufficiently hot to activate proton capture reactions by
magnesium nuclei (notice, anyway, that their results are in qualitatively agreement
with the observations of M13 stars, but no dilution must occur to reproduce the overall Mg 
depletion, while the fit of the magnesium isotopes demands $\sim 30\%$ of pristine matter). 
The proposers of the model, anyway, need two basic, ad hoc, hypotheses to reach a consistent 
result with the GC data:

1) The hypothetical formation of supermassive stars in GCs must be in a limited range of 
initial masses, around 10$^4$\Msun, so that their central temperatures fall in the required 
range 70--80\ MK;

2) The maximum helium enhancement found in SG stars is Y$\simlt$0.4. So it is necessary that 
the core H--burning of such extreme objects ends at a phase when the H content was reduced 
only by $\Delta X \sim 0.15$. The authors suggest that this is justified by the very 
unstable hydrostatic equilibrium of these hypothetical structures.

Apart from these two strong ad hoc choices, 
it is not clear whether these models can explain the $\sim 1$ dex spread in the magnesium 
content observed in metal-poor clusters such as NGC 2419. The trend with the metallicity 
is not obvious in this case, as the temperature at which nuclear activity takes place in 
the central regions is almost unaffected by the metallicity, unlike the HBB conditions
in  massive AGB stars.

\section{Final remarks}
\label{disc}

Among the various species involved in the chemical patterns traced by stars in GCs,
magnesium is the key-element, providing the most valuable information regarding the
nature of the polluters, which released into the intra-cluster medium the gas from which
SG stars formed. This is because unlike other species, such as sodium and oxygen,
magnesium is not affected by any mixing episode connecting the external, convective
regions of the stars with deep layers, where nuclear activity occurred; this holds both
for deep mixing during the red giant branch evolution, which might alter the interpretation of
the observed chemical composition, and also for the second dredge-up in the progenitor 
AGB stars, which changes
the surface chemistry before the thermal pulses phase begins. The magnesium content
reflects the conditions at which the processed gas was exposed, which determine the
extent of the nucleosynthesis experienced. Furthermore, unlike e.g. sodium and
aluminium, the overall magnesium decreases steadily during any p-capture nucleosynthesis
process, which makes the analysis more straightforward, as there is no need to 
consider balance between production and destruction channels, whose trend with the
temperature might be not obvious. Finally, magnesium burning requires higher temperatures
compared to other reactions channels, such as oxygen burning and the activation of the
Ne-Na chain, which restricts the identification of the nature of the possible
pollutors.

Concerning the nuclear processes, in section 3.1 we have discussed the present status of 
uncertainties for the reactions involved in the Mg-Al chains. On the other hand we have 
shown that reproducing the observed isotopic ratios and the overall magnesium spread
observed in clusters of different metallicity requires 3 times higher values of the 
$^{25}{\rm Mg}(p,\gamma)^{26}{\rm Al^{m}}$ reaction, which is larger than the 
uncertainty quoted by LUNA \citep{straniero13}. 

At the temperatures $\sim 100$ MK, relevant for the models studied in this work,
this process is dominated by the resonance at 92 keV. Most of 
the uncertainty in the reaction rate is due to the absence of information about the 
$^{26}$Al level corresponding to this resonance. In particular, a key parameter to 
determine the rate is the ground state feeding factor ($f_0$) for this $^{26}$Al state. 
In fact, as mentioned previously, the $^{25}{\rm Mg}(p,\gamma)^{26}{\rm Al}$ 
resonances decay through complex $\gamma$-ray cascades either to the $5^+$ ground state 
or the $0^+$  isomeric state at E$_X$ =  228 keV. 

For all the resonances at energies 
higher than 92 keV, $f_0$ can be experimentally determined, deducing level branchings 
from $\gamma$-ray detection. This is not possible for the resonance at $92$ keV, because of 
its weakness. For this resonance the value of $f_0$ relies mainly on literature information. 
To our knowledge the main source to get this information is the nuclear compilation 
by \citet{endt88}. Unfortunately, for the 92 keV resonance there is no clear experimental 
information, while for the other levels, also for those at lower energy, i.e. E  = 37 and 
57 keV, the $f_0$, the  determination is well grounded. This is maybe due to the fact that 
literature information in the case of the 92 keV resonance is contradictory. In fact, a 
$f_0$ of about 80$\%$ was deduced in \citet{champagne83a} and \citet{champagne83b}. 
However, the same authors quote a lower value of 61$\%$ in \citet{champagne86}. 
Finally, the compilation by Endt \& Rolfs (1987) gives 85$\%$. The origin of this large 
discrepancy is unknown, it may be possibly attributed to different assumptions on the secondary 
branching ratios. 

In \citet{strieder12} specific primary $\gamma$-ray transitions from 
the 92 keV resonance were not identified, as they used a high efficiency 4$\pi$ BGO 
summing crystal with a limited energy resolution. They adopted a ground-state feeding 
factor of $f_0=60^{+20}_{-10}\%$, taking into account simulations of their experimental 
results and the literature data. The uncertainty of the $f_0$ for this level now sets a 
limit on the reliability of estimates of the $^{25}{\rm Mg}(p,\gamma)^{26}{\rm Al}$ 
reaction in the critical temperature range.

This paper shows that measuring again this feeding factor is  important to reduce the 
uncertainty in this cross section, relevant to the issue of the magnesium and isotopic 
ratios  observed in the second generation of GCs.


\end{document}